\documentclass[12pt,preprint]{aastex}

\slugcomment{Not to appear in Nonlearned J., 45.}

\shorttitle{Three  Li-Rich K giants}

\begin{document}

\title{Three  Li-rich K giants: IRAS~12327-6523, IRAS~13539-4153, and IRAS~17596-3952}


\author{Bacham E. Reddy\altaffilmark{1}
\affil{Indian Institute of Astrophysics, Bangalore-560034, INDIA}
\and
David L. Lambert}
\affil{The W.J. McDonald Observatory, The University of Texas at Austin, Austin, TX-78731}
\email{ereddy@iiap.res.in  dll@astro.as.utexas.edu}


\altaffiltext{1}{Visiting Astronomer, Cerro Tololo Inter-American Observatory.
CTIO is operated by AURA, Inc.\ under contract to the National Science
Foundation.}

\begin{abstract}
We report on  spectroscopic analyses of three K giants previously suggested to be
Li-rich:
IRAS~12327-6523, IRAS~13539-4153, and IRAS~17596-3952.
High-resolution optical spectra and the LTE model atmospheres are used to derive
the stellar parameters: ($T_{\rm eff}$, log $g$, [Fe/H]), elemental abundances,
and the
isotopic ratio $^{12}$C/$^{13}$C.
IRAS~13539-4153 shows an extremely high Li abundance of
$\log\epsilon$(Li) $\approx$ 4.2, a value ten
times more than the present Li abundance in the local interstellar medium.
This is the third highest Li abundance yet reported for a
K giant.
IRAS~12327-6523
shows a  Li abundances of $\log\epsilon$(Li)$\approx$ 1.4.
IRAS~17596-3952 is a rapidly rotating ($V{\sin i}$ $\approx$ 35 km s$^{-1}$)
K giant with $\log\epsilon$(Li) $\approx$ 2.2.
Infrared  photometry which
shows the  presence of an IR excess suggesting mass-loss. 
A comparison is made between these three stars and previously
recognized Li-rich giants.
\end{abstract}


\section{Introduction}

The maximum lithium abundance of main sequence
stars of near-solar metallicity spans a small range: $\log\epsilon$(Li) $\approx$ 3.3 for
[Fe/H] $\approx$ 0.0 to 3.1 for [Fe/H] $\approx$ $-$0.3 (Lambert \& Reddy 2003). This
lithium abundance is identified with that of the interstellar gas from which the
stars formed.
Lithium is predicted to be destroyed throughout a main sequence star except in the outermost
layers (1\% -- 2\% by mass). Even in this thin skin which includes the atmosphere,
lithium may be depleted by processes not yet fully understood. The Sun, for example
has a lithium abundance about a  factor of 100 less than the meteoritic abundance.

When the main sequence star evolves to become a red giant, the convective
envelope grows and dilutes the lithium such that its surface abundance is
greatly reduced. Iben (1967a,b) predicted that  giants of approximately solar
metallicity would experience  dilution  by a factor of 1.8 dex at 3$M_{\rm \odot}$ and
1.5 dex at 1$M_{\odot}$. This dilution occurs at the onset of the first dredge-up
resulting from growth of the convective envelope
which also results in predicted changes of the surface (elemental and isotopic)
abundances of helium, beryllium,
boron, carbon, nitrogen, and oxygen.
Adopting the predicted lithium dilution and the maximum observed lithium
abundance of main sequence stars, lithium in red giants is expected not
to exceed $\log$$\epsilon$(Li) $\approx$ 1.5.
 A red giant star that has evolved through the first dredge-up
but has a surface abundance
greater than
$\log$$\epsilon$(Li) $\approx$ 1.5
is declared to be `Li-rich'.

Lithium-rich K giants are rare.
Since the  discovery of the first Li-rich K giant,  HD~112127, two decades ago (Wallerstein \& Sneden 1982), the
collection of Li-rich stars has grown to only about
three dozen. Many like HD~112127
were discovered serendipitously. Brown et al. (1989) undertook a systematic
search for Li-rich stars among K and late G-type giants.
They found at most  ten Li-rich  giants
out of the 670 stars observed.
Another important survey known as ``Pico Dias Survey" (PDS) by Gregorio-Hetem et al. (1992)
found, as a byproduct, potentially Li-rich giants
(de la Reza et al. 1997). A few of these PDS candidate Li-rich K giants
have been spectroscopically studied to determine their lithium
abundance (e.g; Reddy et al. 2002a; Drake et al. 2002).

Here, we report results from the first  high-resolution spectroscopic analyses of
three Li-rich candidates taken from de la Reza et al. (1997) who
proposed the stars  as probably Li-rich from the observation that
the strength of the  Li\,{\sc i} 6707 \AA\ doublet was
equal to or greater than the strength of the   Ca\,{\sc i}  6718~\AA\ line.
It turns out that the lithium abundances of the three stars are quite different.
The most extreme example is IRAS 13539-4153 (also known as
PDS 68) with $\log\epsilon$(Li) $\approx$4.1, an abundance almost 1 dex higher than the
star's likely initial lithium abundance.
The second star IRAS 17596-3952 has $\log$$\epsilon$(Li) $\approx$ 2.2 and is most
probably a Li-rich giant, i.e., a first dredge-up giant with lithium
in excess of the  limit $\log$$\epsilon$(Li) $\approx$ 1.5.
The third star IRAS 12327-6523 has $\log$$\epsilon$(Li) $\approx$ 1.4 and so
is not a certain Li-rich star; it is  possibly a post-first dredge-up
star with lithium depleted to the predicted level.
Following presentation and  discussion of the abundance  analysis, we comment
on the similarities between  the new and previously  known Li-rich  giants.

\section{Observations}

Spectra of the stars listed in Table 1
 were obtained with the 4~m Blanco telescope and the Cassegrain
echelle spectograph
at the Cerro Tololo Inter-American Observatory in Chile.
In addition to obtaining spectra of the three Li-rich candidates,
we obtained a high-quality spectrum of HD 19745, a Li-rich giant
analysed previously by de la Reza \& da Silva (1995).

The spectrograph was set to
cover  the wavelength interval 5000~\AA\ to 8200~\AA\ without gaps at a  resolving
power of $R\approx$ 35,000. Each star was exposed
thrice with a total integration time ranging from 60~min to 90~min.
Raw two-dimensional spectra were reduced to one-dimensional
wavelength-calibrated spectra using appropriate reduction tools
available in the  $IRAF$ package.
A Th-Ar hollow cathode spectrum provided the wavelength calibration.
In each spectrum of a Li-rich candidate star, the signal-to-noise at 6707 \AA\
is  between 100 and 200.

\section{Analysis}

\subsection{Stellar Properties}

Stellar parameters $T_{\rm eff}$, ${\rm log \rm g}$, and [Fe/H])
were obtained from the spectra by a standard procedure
using  $ATLAS$ model atmospheres computed using the
no convective overshoot option (Kurucz 1994).
The effective temperature ($T_{\rm eff}$)
was obtained from   the requirement that all Fe\,{\sc i} lines
give the  same abundance independent of the lower state's excitation potential (LEP).
About  40 Fe\,{\sc i} lines spanning the range LEPs of 1 to 5 eV
 were measured for IRAS 12327-6523 and IRAS 13539-4153 but fewer lines
were measureable  for the rapidly- rotating
K giant IRAS~17596-3952.
Lines are weak to  moderately strong ($W_{\lambda}$ $\leq$ 150~m\AA).
Surface gravity (${\rm log \rm g}$) was obtained with the condition
that the Fe\,{\sc i} and Fe\,{\sc ii} lines give the same abundance. The microturbulent
velocity ($\xi_{\rm t}$) was determined by requiring that the Fe\,{\sc i} lines
and Ni\,{\sc i} lines
give the same (Fe or Ni) abundance irrespective of their equivalent width.
The parameters were determined by an iterative process using models selected from
the $ATLAS$ grid.
Adopted parameters are given in Table~1. Our results for HD~19745
are in good agreement with the previous determination by Castilho et al. (2000).

A full abundance analysis was made with the selected model atmospheres.
Lithium  abundance and the $^{12}$C/$^{13}$C ratio are discussed
below. Seventeen elements from Na to Nd in addition to Fe were
included in the analysis. Given that [Fe/H] $\simeq 0$ for the
stars which probably belong to the thin disk, one expects [X/Fe] $\simeq 0$
for the 17 elements. This is what is found to within the measurement
errors, except for the suggestion that [Ca/Fe] $\simeq -0.3$ for IRAS 12327-6523
and IRAS 13539-4153. The detailed analysis was not attempted for the rapidly
rotating IRAS 17596-3952. 

Inspection of the spectra shows two unusual aspects worthy of comment:
unusually broad lines of IRAS~17596-3952, and an asymmetric
H$\alpha$ profile of IRAS~12327-6523.

The lines of IRAS 17596-3952 are obviously broader than those of the
other two stars which have line widths typical of K giants.
 The additional broadening is attributed to a more rapid than average
rotational velocity.
 To determine the projected rotational velocity
($V{\rm sin i}$), we match the observed spectrum to a synthetic spectrum
computed with the following broadening parameters:
the
velocity $V{\rm sin i}$, a macroturbulent velocity $V_{\rm t}$, and the instrumental profile width.
 The value  $V_{\rm t}$ = 3~km s$^{-1}$ was adopted, as found to be
typical of K giants (Gray 1989).
The instrumental profile width was taken as Gaussian with a
$FWHM$ = 0.21~\AA\ at 6707~\AA\ as measured from the Th emission lines in
the hollow cathode spectrum.
The non-Gaussian form (Gray 1976) of the broadening resulting from
solid body stellar rotation is incorporated in the line
formation code MOOG (Sneden 1973).
Unblended  Fe\,{\sc i} lines
at 6703.5~\AA, 6705.1~\AA\ and 6170~\AA\ were selected for
the estimation of $V{\rm sin i}$.
A $\chi^{2}$-test was used to select the best value of $V{\rm sin i}$.
The mean value  from the fits to the three   Fe\,{\sc i} lines
is given in Table~1 for each star.

Profiles of the H$\alpha$ line in the four stars and Arcturus are shown in
Figure~1. It is clear that IRAS~12327-6523's H$\alpha$ line is distinctly
asymmetric with the line core shifted to the blue relative to the
photospheric lines. An asymmetric H$\alpha$ line is seen in some
other Li-rich giants (Drake et al. 2002).

\subsection{Lithium Abundance}

An immediate indication that the stars might be Li-rich was found at the
telescope from the obvious presence of a strong absorption line
attributable to the Li\,{\sc i} resonance doublet at 6707.8~\AA.
Closer inspection reveals also the presence of the excited Li\,{\sc i}
line at 6103.6~\AA\ in the spectra
of IRAS~13539-4153 and
HD~19745. Lithium abundances were found
by spectrum synthesis.

Synthesis of the 6707 \AA\ feature used the
critically examined line list in the vicinity of the resonance line
taken from Reddy et al. (2002b) with the component structure of the
Li\,{\sc i} resonance doublet taken
from Hobbs, Thorburn, \& Rebull (1999).
Lithium is assumed to be purely $^7$Li.
The line list for 10~\AA\ centered  on the Li\,{\sc i}
line $\lambda$ 6707~\AA\  was tested
by successfully reproducing the ultra-high ($R$ $\approx$ 150,000) resolution
spectrum (Hinkle et al. 2000) of the very Li-poor K-giant Arcturus 
(see for details: Reddy et al. (2002a)).
Then, synthetic spectra were computed and matched to the  spectrum
of each program star
by adjusting the lithium abundance to fit the 6707.8~\AA\
feature.
Comparisons of synthetic and observed spectra
are shown in Figure~2 \& 3.
Lithium abundances derived from the LTE analysis are given in Table~2.
Uncertainties in the derived lithium abundances (Table~2)
are estimated from uncertainties in the atmospheric  parameters:
$\delta$$T_{\rm eff}$ = 100~K,
$\delta {\rm log \rm g}$ = 0.5, $\delta$$\xi_{t}$ = 0.5 km s$^{-1}$, and $\delta$[M/H] = 0.25~dex
lead uncertainties of 0.25, 0.05, 0.05, and 0.03 dex, respectively.

For IRAS~13539-4153 and HD~19745, the lithium abundance
has been obtained also using the subordinate Li\,{\sc i}
line at 6103~\AA. The line
list for this region was compiled using Kurucz's line list (Kurucz 1994).
In Figure~4, we compare synthetic and observed spectra. 
The lithium abundances (Table~2)
 derived from the 6707~\AA\ and 6103~\AA\ lines
are in good agreement for HD~19745, and in good agreement too
with previously published results by
de la Reza \& da Silva (1995).

In Figure~2, we  draw attention to a failure to fit the
Li\,{\sc i} 6707 \AA\ profile of IRAS~13539-4153.
The line  is broader for its depth than calculated:
the abundance  $\log\epsilon$(Li) = 2.35 fits  the core but
 $\log\epsilon$(Li) = 4.1 is required to fit the wings, but
this abundance produces a central depth greater than
observed. The 6103\AA\ Li\,{\sc i} line
is matched by the abundance
$\log\epsilon$(Li) = 4.1 and the $V{\sin i}$ of the metal lines. 
We suppose that this latter abundance
is the stellar abundance and that the
6707~\AA\
line's core is affected by emission due to chromospheric activity
or by non-LTE effects.
It is unclear
why IRAS~13539-4153's 6707~\AA\ line is so affected but HD~19745's line
of comparable strength is not.
(IRAS~13539-4153's  6707~\AA\ profile is well fit
by increasing the $V{\rm sin i}$ to 16 km s$^{-1}$ but this
broadening fits no other lines in the spectrum (see Figure~2).)

An additional
uncertainty arises from our assumption of LTE for line
formation.
For  the super Li-rich giants HD~19745 and IRAS~13539-4151
non-LTE corrections  indicate a correction of
0.1~dex upward for the 6103~\AA\ line and  0.2~dex
downward for the  6707~\AA\ line  (Carlsson et al. 1994).
For the less lithium-rich stars
IRAS~12327-6523 and IRAS~17596-3952, the non-LTE  correction is
around 0.1~dex upward for the 6707~\AA\ line. 
Since these corrections are small
and subject to their uncertainties, we include them as part of the
overall abundance uncertainty. (The non-LTE abundances are also given in Table~2.)
The total
error given in Table~2 is the quadratic sum of all the errors
discussed above.

\subsection{The $^{12}$C/$^{13}$C ratio}

Standard models make predictions about the  first dredge-up's effects on
the surface abundances of Li, C, N, and O.
Observations of normal red giants confirm these predictions,
qualitatively, if not always quantitatively.
Interpretation of lithium abundances is assisted and possibly
complicated by knowledge of the C, N, and O abundances.
Here, we restrict the analysis to the carbon isotopic ratio $^{12}$C/$^{13}$C. 
One expects an unevolved star to have a ratio close to $^{12}$C/$^{13}$C $\approx$ 90, the solar
photospheric value.
A lower ratio ($\approx$ 20) is predicted for giants after the first dredge-up. Some
giants show even lower $^{12}$C/$^{13}$C ratio.

The $^{12}$C/$^{13}$C ratio was extracted from
a complex of $^{13}$CN lines at 8005.7~\AA\ and adjacent $^{12}$CN lines (Figure~5).
The ratio $^{12}$C/$^{13}$C is determined by spectrum synthesis assuming
solar abundances (Lodders 2003) for C and N.
For HD~19745  we obtained $^{12}$C/$^{13}$C = 16$\pm$2
which is in good agreement with the ratio of 15
derived by da Silva, de la Reza, \& Barbuy (1995). For IRAS 12327-6523, we
obtain
$^{12}$C/$^{13}$C = 6$\pm$1. A limit of $^{12}$C/$^{13}$C $\geq$ 20 was found for
IRAS~13539-4153. Spectral lines of IRAS~17596-3952 are too smeared, due to its high rotation,
to estimate the ratio. The low $^{12}$C/$^{13}$C ratios confirm that the Li-rich
giants have experienced the first dredge-up.

\section{The Li-rich clans}

When considered with the population of field K giants, our
trio stand apart by reason of their high lithium abundance.
That is why they were selected for an abundance
analysis.
The question we consider here is how the trio fit in with
the previously analyzed Li-rich giants.

Charbonnel \& Balachandran (2000) took advantage of {\it Hipparcos}
parallaxes to estimate the absolute luminosity of Li-rich and
Li-normal red giants. They suggested that the Li-rich giants
could be divided into three clans: warm giants for which lithium
dilution has not been completed; giants at the bump in the luminosity
function; giants in the early phases of asymptotic giant branch (AGB)
evolution -- see Figure 6.

Our trio (and HD~19745) are not in the {\it Hipparcos}
catalogue. Therefore, we estimate their luminosities using
the spectroscopically-derived ${\rm log \rm g}$ and $T_{\rm eff}$ and
computed evolutionary tracks
for
different stellar masses and metallicities (Girardi et al. 2000).
By interpolation, we found the  track (i.e., stellar mass) for which
the ${\rm log \rm g}$ and $T_{\rm eff}$ best agreed with
the observed log $g$ and $T_{\rm eff}$. The luminosity is then read
from the track.
Estimated  luminosities and masses are given in Table~1. The estimated
errors due to uncertainties in model parameters ($\delta T_{\rm eff}$ = 100~K,
$\delta {\rm log \rm g}$ = 0.25, and $\delta [Fe/H]$ = 0.2) are shown in Table~1.
According to their luminosity and effective temperature
the three stars and HD~19745  have completed the dilution
associated with the first dredge-up. The stars are shown in Figure~6 with
others from our earlier studies. 
For stars taken from earlier
studies uncertainties of $\delta T_{\rm eff}$ = 100 K and $\delta {\rm log \rm g}$ = 0.25 are
assumed (Figure~6).

The two most Li-rich giants
IRAS~13539-4153 and IRAS 17596-3952 belong to
the `bump' clan.  By the spectroscopically estimated luminosity,
we place HD~19745 also among this clan.
The range of lithium abundances, $^{12}$C/$^{13}$C ratios, and
$V{\rm sin i}$ covered by these three stars are spanned by Charbonnel \&
Balachandran's  six stars: 
IRAS~17596-3952's $V{\rm sin i}$ with 35 km s$^{-1}$ extends the
upper limit previously set at
25 km s$^{-1}$.

IRAS12327-6523 is assigned to
Charbonnel \& Balachandran's `early-AGB' clan with its original
five members with lithium abundances ranging from
$\log\epsilon$(Li) = 1.5 to 2.8. IRAS~12327-6523 is now the least
lithium rich of the clan. Strictly speaking, the star is
not lithium rich because standard dilution by the first dredge-up
could have left this relatively massive star with the
now observed abundance. However, standard models of the first dredge-up
cannot account for the low $^{13}$C/$^{12}$C (=6) ratio;
a ratio of 20 - 30 is predicted. It
remains to be shown that the lower than expected carbon isotopic ratio
can be achieved by modifications to standard models without severe
reduction of the surface lithium abundance.

Greg\'{o}rio-Hetem et al. (1992, also Greg\'{o}rio-Hetem, Castilho,
\& Barbuy 1993) found that the great
 majority of the Li-rich
giants possess an infrared excess. Drake et al. (2002) noted that
 every Li-rich
rapid rotator in a sample of 20 had an infrared excess. Far-IR
fluxes from the {\it IRAS} catalog show that our three stars
(and HD 19745) have a strong excess at 12 $\mu$m and longer wavelengths,
as expected from their designation as an IRAS source.
Fluxes estimated from the {\it 2MASS} J, H, and K magnitudes 
show that the IR excess does not extend to these shorter
wavelengths. Our three Li-rich stars, therefore, continue the
correlation between Li-richness and IR excess, and, in particular,
IRAS 17596-3952 strengthens the impression that all Li-rich
rapid rotators have a far IR excess.

\section{Concluding remarks}

We have performed the first detailed analysis of three
Li-rich candidate K giants and a reanalysis of  HD~19745.
Two candidates are confirmed as certainly Li-rich:  IRAS~13539-4153 is among the most
Li-rich giants yet analysed. The status of IRAS~12327-6523 as Li-rich
might be questioned.
Three of the four giants --  HD~19745, IRAS~13539-4153, and IRAS~17596-3952 --
 belong to RGB-clump  collection of Li-rich giants
and IRAS~12327-6523 belongs to the early-AGB collection in the H-R diagram.
This new trio and the other Li-rich giants analyzed previously by us
confirm that  characteristics of the RGB-bump clan that set them apart from
other red giants: lithium abundance as high 
as $\log(Li) \simeq 4$, a far-infrared excess, and in the mean a high
rotational velocity. 

Charbonnel \& Balachandran (2000) linked the RGB-bump clan to
the RGB bump where evolution along the giant branch is
slowed when the H-burning shell burns through the molecular weight
discontinuity left from the initial growth of the convective envelope
which provided the first dredge-up. (The Li-rich giants at higher
luminosity and of higher mass were associated with a molecular
weight discontinuity predicted to occur in early-AGB stars 
as a result of the second dredge-up.). Erasure of the molecular weight
gradient makes it possible for `extra' mixing to occur between the
convective envelope and the top of the H-burning shell. Mixing results
in the conversion of $^{3}$He to $^7$Be and then $^7$Li by the
Cameron-Fowler (1971) mechanism. 
This extra mixing may be internally or externally triggered.
A proposed trigger  should explain the concentration
in luminosity of the Li-rich giants, the tendency for Li-rich
giants to be rapid-rotators and the association with a far infrared-excess.
A reader challenged to take up the search for the trigger might
profitably read the proposals advanced by
 Palacios, Charbonnel, \& Forestini (2001) and
Denissenkov \& Herwig (2004).

\acknowledgments

David L. Lambert acknowledges the support of the Robert A. Welch
Foundation of Houston, Texas.
We would like to acknowledge with thanks
use of 2MASS and NASA ADS data bases.

\clearpage

\begin{deluxetable}{crrrrrrcrr}
\tabletypesize{\scriptsize}
\tablecaption{Properties of the four Li-rich K giants. \label{tbl-1}}
\tablewidth{0pt}
\tablehead{
\colhead{Star} & \colhead{$T_{\rm eff}$}   & \colhead{log $g$}   &
\colhead{$\xi_{t}$} &\colhead{$V_{\rm helio}$} & \colhead{[Fe/H]$^{a}$} & \colhead{$V{\rm sini}$} &\colhead{W$_{\lambda}$($\lambda$6707)}
&\colhead{M/M$_{\odot}$}& \colhead{L/L$_{\odot}$} \\
\colhead{} & \colhead{K}   & \colhead{cm s$^{-2}$}   &
\colhead{km s$^{-1}$} &\colhead{km s$^{-1}$} & \colhead{} & \colhead{km s$^{-1}$} &\colhead{m\AA)}
&\colhead{}& \colhead{}
}
\startdata
HD19745$^{b}$           & 4700$\pm$100 & 2.25$\pm$0.25 & 1.5$\pm$0.2 &$-$15.4$\pm$1.0 &$-$0.05 & 3.0$\pm$1.0& 470 & 2.2$\pm$0.6 & 1.9  \\
IRAS~12327$-$6523 & 4200$\pm$100 & 1.25$\pm$0.25 & 2.0$\pm$0.2 &$-$18.9$\pm$1.0 &$-$0.40 & 7.5$\pm$1.0& 310 & 2$\pm$1.0   & 2.9\\
IRAS~13539$-$4153 & 4300$\pm$100 & 2.25$\pm$0.25 & 2.0$\pm$0.2 &$-$87.3$\pm$1.3 &$-$0.13 & 5.0$\pm$1.0& 630 & 0.8$\pm$0.7 & 1.6\\
IRAS~17596$-$3952 & 4600$\pm$100 & 2.50$\pm$0.50 & 1.9$\pm$0.2 &$-$27.2$\pm$2.3 &$+$0.10 & 35.0$\pm$ 5& 300 & 1.7$\pm$0.6 & 1.7\\
\enddata
\tablenotetext{a}{ [Fe/H] = log $\epsilon$(Fe) - log $\epsilon$(Fe)$_{\odot}$ with the solar iron abundance 
                   of log $\epsilon$(Fe) = 7.47 taken from Lodders (2003) }
\tablenotetext{b}{ Castilho et al. (2000) obtained $T_{\rm eff}$ = 4750~K, 
{\rm log \rm g}= 2.5, $\xi_{t}$ = 1.2 and [Fe/H] = 0.0.}
\end{deluxetable}

\clearpage

\begin{deluxetable}{crrrrrrrr}
\tabletypesize{\scriptsize}
\tablecaption{Derived Li abundances and carbon istopic ratios.
\label{tbl-1}}
\tablewidth{0pt}
\tablehead{
\colhead{Abundance \& Feature} & \colhead{IRAS~12327-6523}   & \colhead{IRAS~13539-4153}   &
\colhead{IRAS~17596-3952} & \colhead{HD19745}
}
\startdata
LTE log $\epsilon$(Li) (LiI $\lambda$ 6103) & ...      & 4.15$\pm$0.3      &  ...   & 3.90$\pm$0.3   \\
NLTE log $\epsilon$(Li) (LiI $\lambda$ 6103) & ...      &4.2      &  ...   & 4.0  \\
                                            &           &         &        &       \\
LTE log $\epsilon$(Li) (LiI $\lambda$ 6707) & 1.40$\pm$0.2      & 4.10$\pm$0.3      & 2.2$\pm$0.2      & 3.70$\pm$0.3   \\
NLTE log $\epsilon$(Li) (LiI $\lambda$ 6707) & 1.6      & 3.9      & 2.3      & 3.4   \\
                                            &           &         &        &       \\
$^{12}$C/$^{13}$C                       & 6.0$\pm$1 & $\geq$20  &  ...      & 16$\pm$2.0 \\
\enddata

\end{deluxetable}
\clearpage

\begin{figure}
\plotone{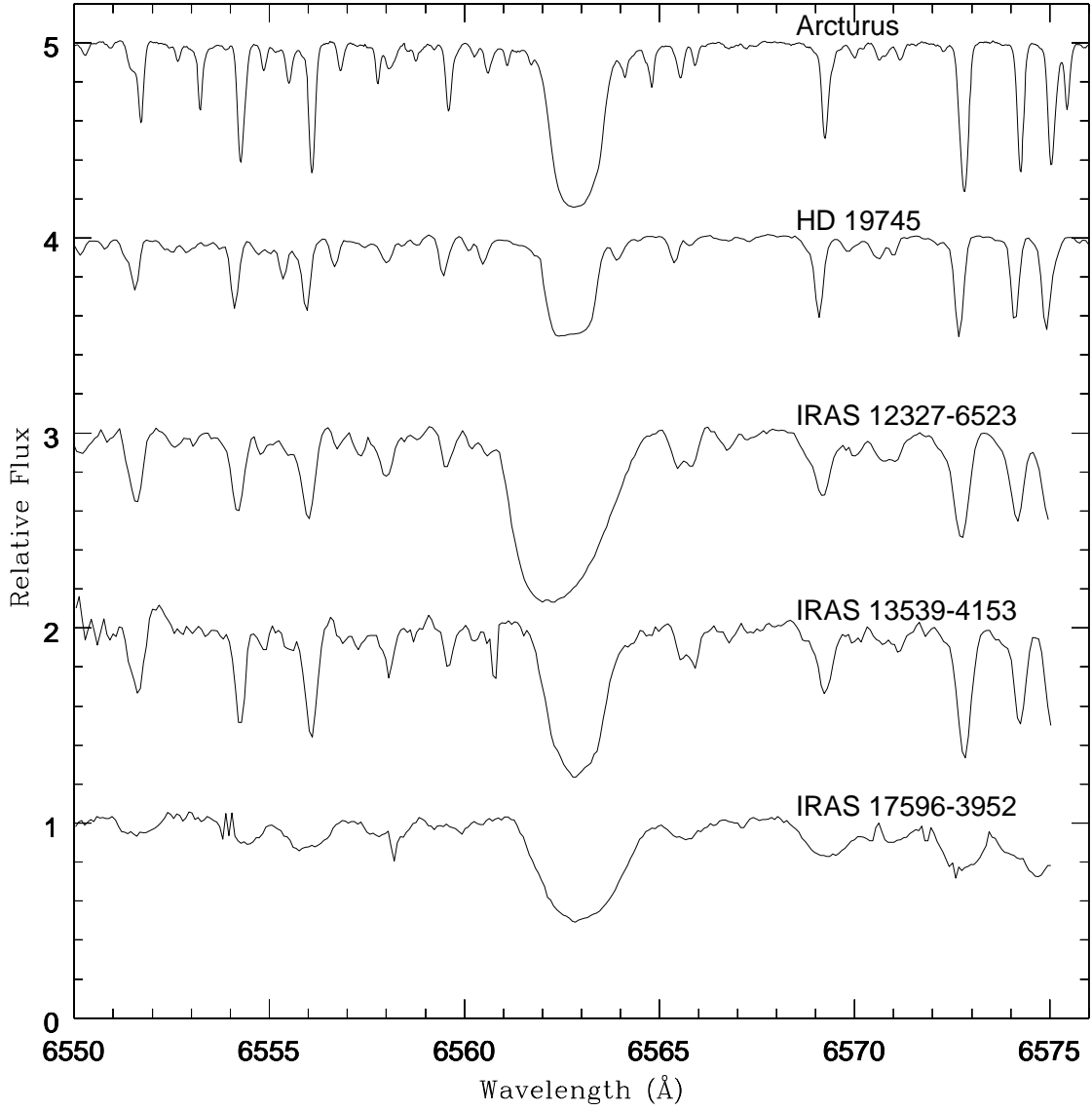}
\caption{Spectra of Li-rich K giants and Arcturus near the H$\alpha$ line. 
Note the asymmetric H$\alpha$ profile of IRAS~12327-6523}
\end{figure}

\clearpage

\clearpage

\begin{figure}
\plotone{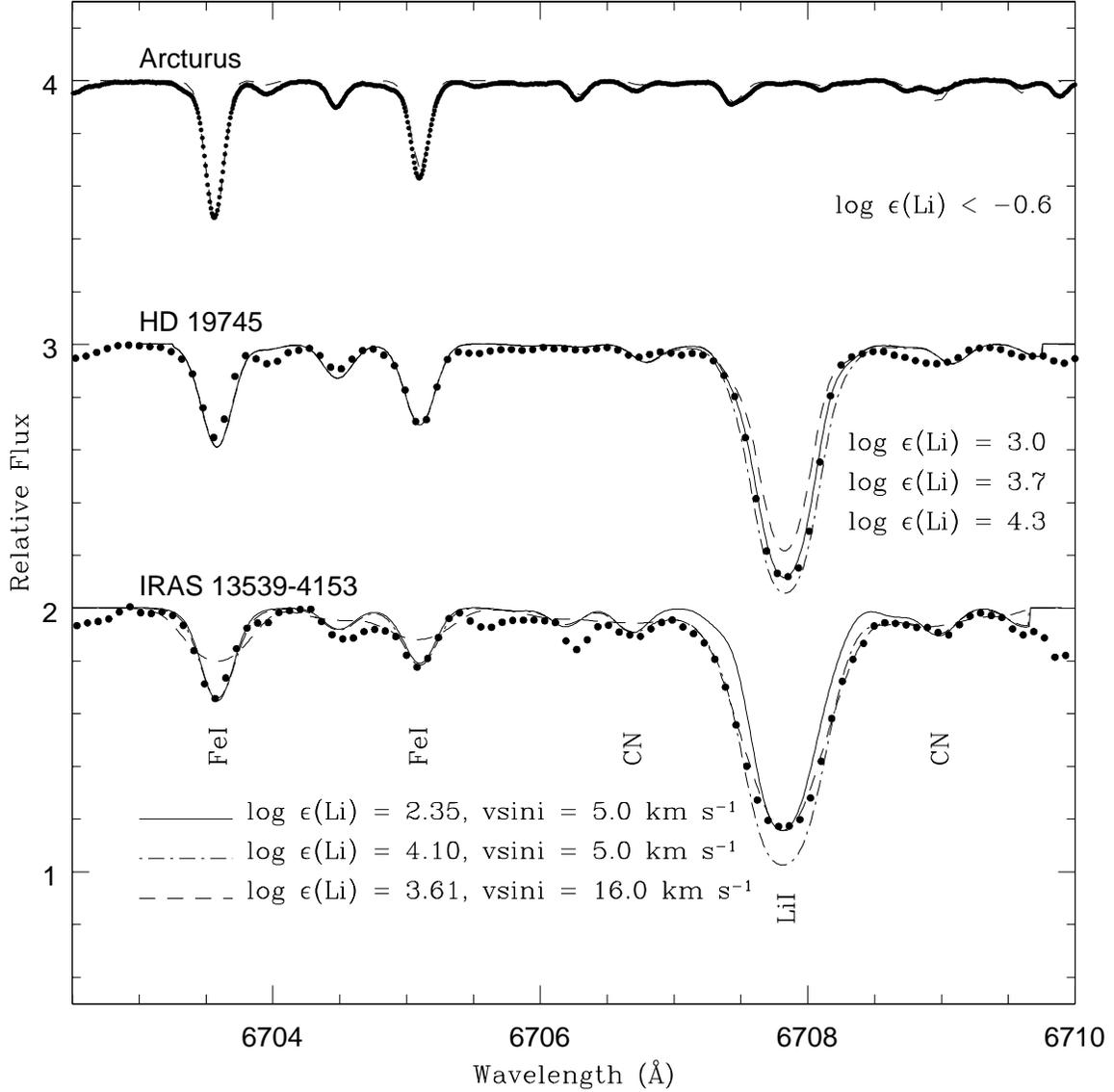}
\caption{Observed (filled circles) and synthetic spectra (lines) of a region around
the Li~I line at 6707.6~\AA\ for Arcturus, HD~19745, and IRAS~13539-4153. Note that the very strong LiI line at 6707.8\AA\
in HD19745 and IRAS13539 and absence of the same line in Arcturus. 
The stars have a similar effective temperature. The synthetic spectra with log $\epsilon$(Li) $<$ $-$0.6 and 3.7
provide the best fit for Arcturus and HD~19745, respectively. See text for discussion of synthetic spectra for IRAS~13539-4153.
\label{fig2}}
\end{figure}

\clearpage
\begin{figure}
\plotone{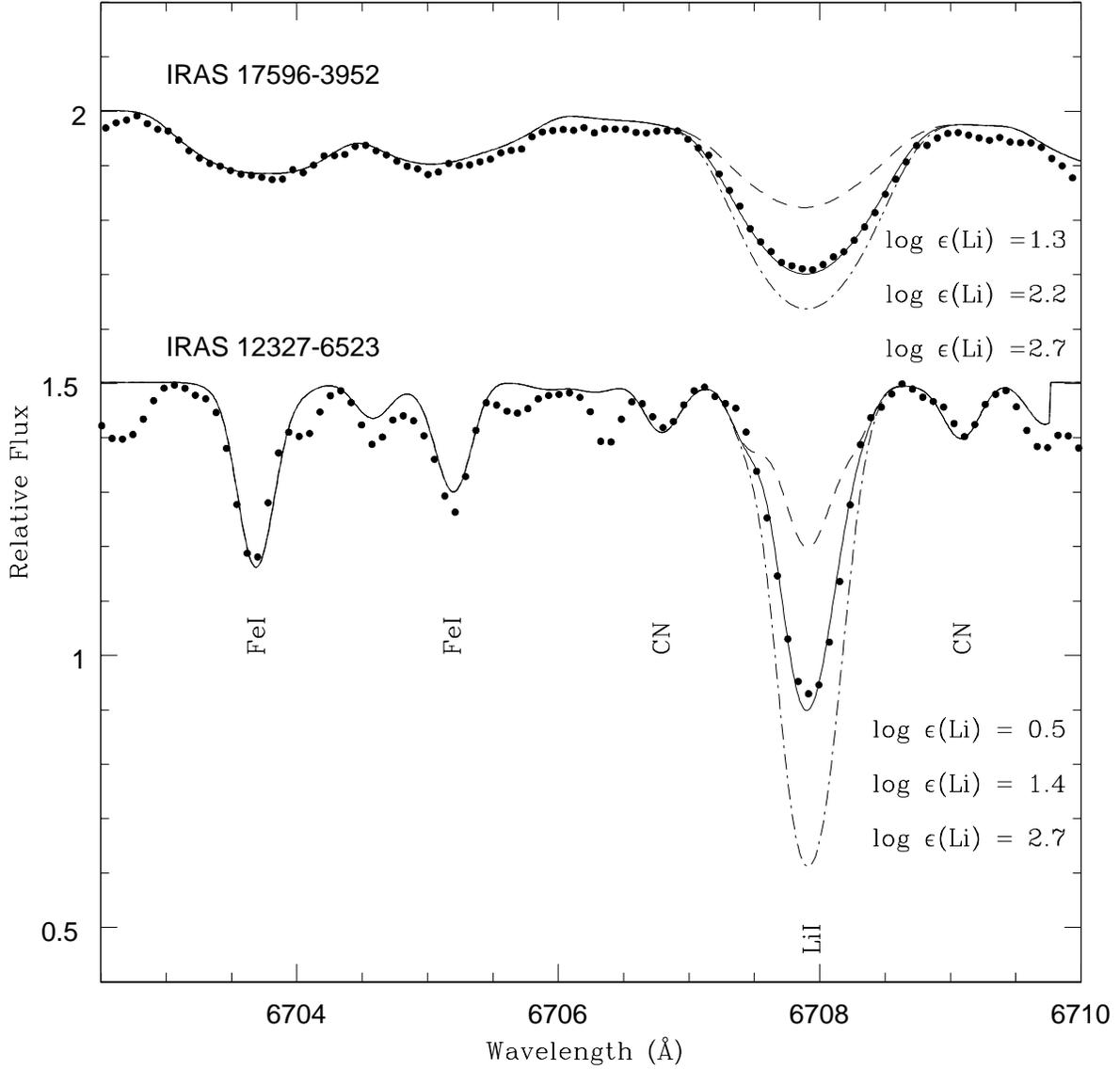}
\caption{Observed (filled circles) and synthetic spectra (lines) of a region around
the Li~I line at 6707~\AA\ for IRAS~17596-3952 and IRAS~12327-6523.
Spectra at 6707~\AA\ were predicted
for the three different Li abundances as noted in the figure.
The solid line shows the best fit to the observations: log $\epsilon$(Li) = 2.0 for IRAS~17596-3952 and 1.4 for IRAS~12327-6523.
\label{fig3}}
\end{figure}
\clearpage

\begin{figure}
\plotone{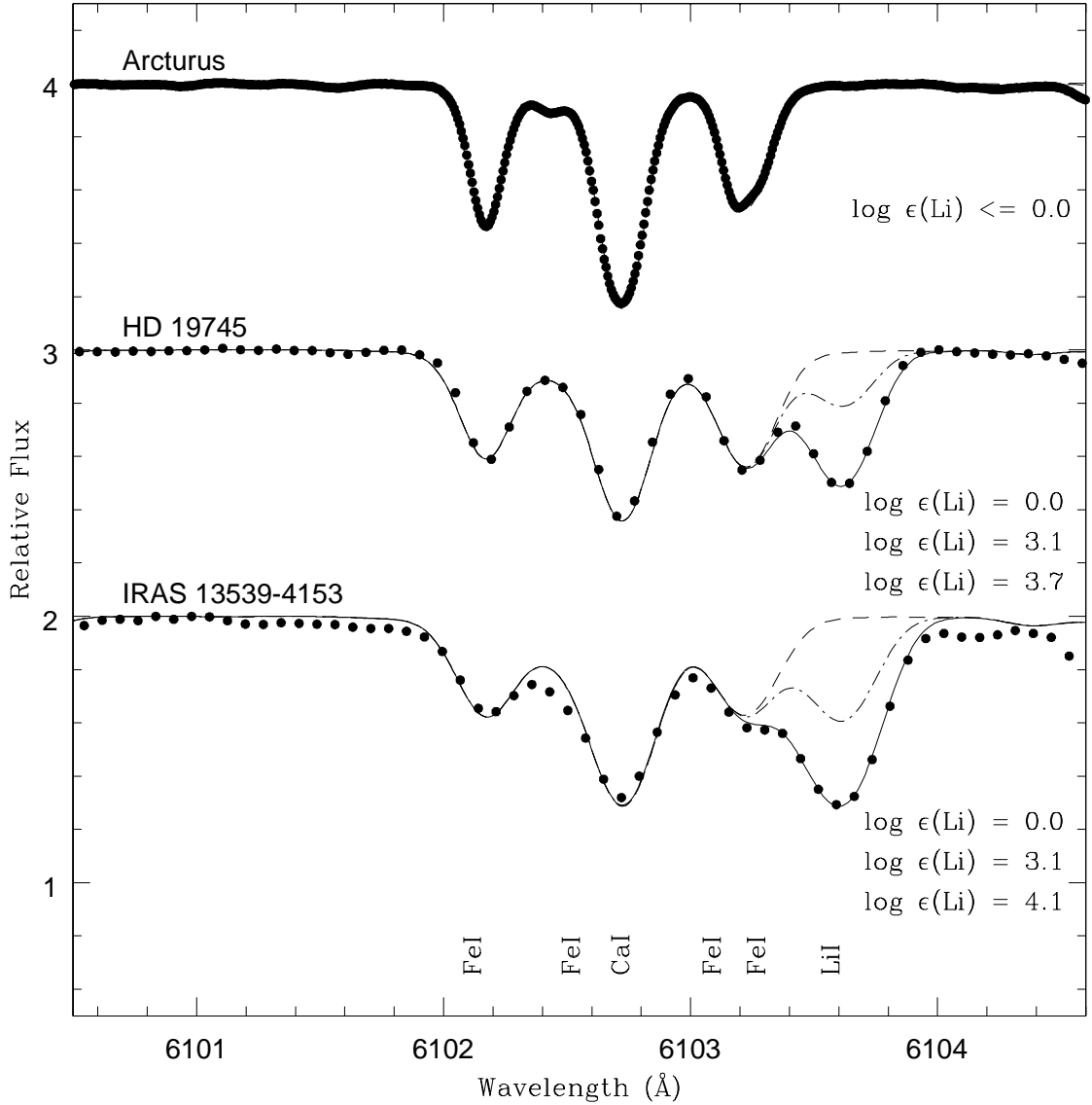}
\caption{Observed (filled circles) and synthetic spectra (lines) of a region around
the Li~I line at 6103.5~\AA\ for Arcturus, HD~19745, and IRAS~13539-4153.
Spectra at 6103~\AA\ were predicted
for the three different Li abundances as noted in the figure.
The solid line shows the best fit to the observations: log $\epsilon$(Li) $<$ $-$0.6, 3.7, and 4.1 for Arcturus, HD~19745, and
IRAS~13539-4153, respectively.
\label{fig4}}
\end{figure}

\clearpage
\begin{figure}
\plotone{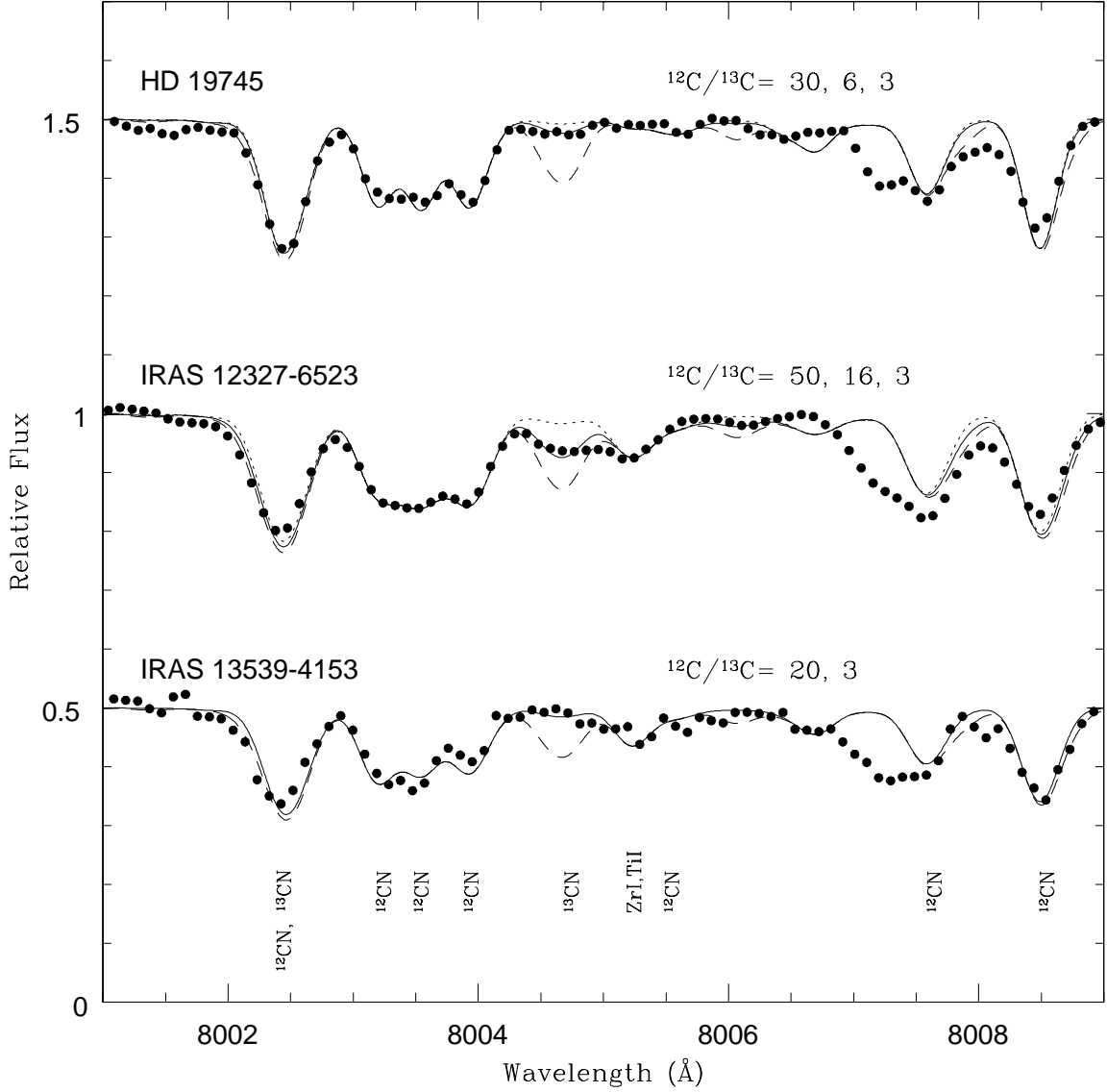}
\caption{Determination of the $^{12}$C/$^{13}$C ratio
for three Li-rich K giants. In all the cases
solid line is the best predicted spectrum for $^{13}$CN line at 8004.8~\AA\ resulting
$^{12}$C/$^{13}$C = 6, 16 and $\geq$20 for HD~19745, IRAS~12327$-$6523, and IRAS~13539$-$4153, respectively.}
\label{fig6}
\end{figure}

\clearpage

\begin{figure}
\plotone{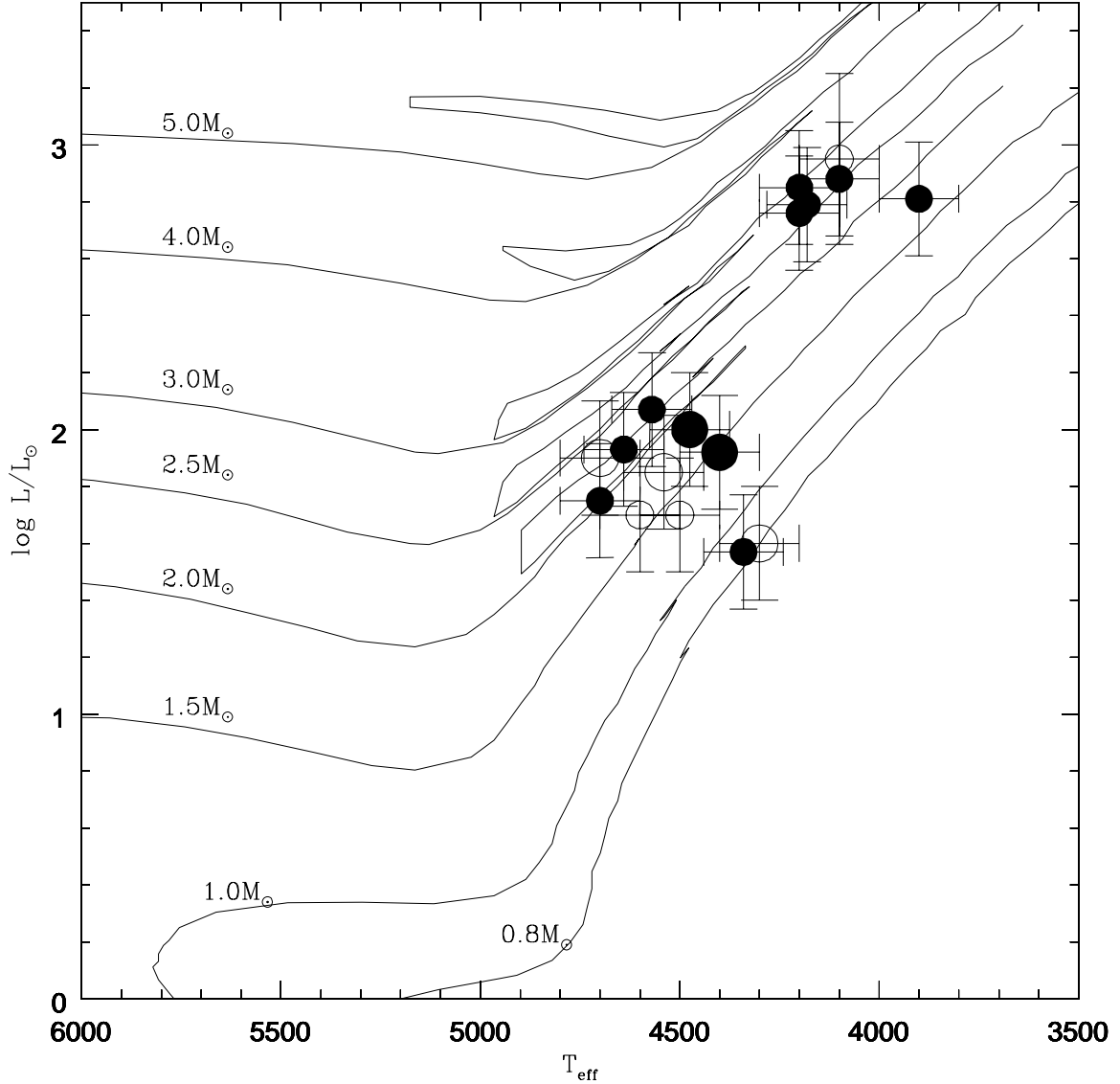}
\caption{Li-rich giants in H-R diagram. Li-rich giants at the early-AGB phase
(log L/L$_{\odot}$ = 2.8) and the luminosity function bump (log L/L$_{\odot}$ = 1.8)
from 
Charbonnel \& Balachandran (2000) are shown by solid circles.
The four Li-rich giants including HD~19745 from this study
and one each from Reddy et al. (2002a) and Drake et al. (2002) are
shown by open circles. Bigger
symbols represent stars with log $\epsilon$ (Li) $\geq$ 3.3 and the smaller symbols represent
stars with log $\epsilon$ (Li) = 1.4 to 3.0. Evolutionary tracks of [Fe/H] = 0 
(Girardi et al. 2000) for various
masses are shown.} 
\end{figure}

\end{document}